\documentstyle[12pt]{article}
\begin{document}
\def\thefootnote{\fnsymbol{footnote}}
\begin{flushright}
KANAZAWA-00-12  \\ 
December, 2000\\
\end{flushright}
\vspace*{2cm}
\begin{center}
{\LARGE\bf Neutrino mass due to the mixing among neutrinos and gauginos}\\
\vspace{1 cm}
{\Large  Daijiro Suematsu}
\footnote[1]{e-mail:suematsu@hep.s.kanazawa-u.ac.jp}
\vspace {0.7cm}\\
$^\ast${\it Institute for Theoretical Physics, Kanazawa University,\\
        Kanazawa 920-1192, JAPAN}
\end{center}
\vspace{2cm}
{\Large\bf Abstract}\\  
We consider the explanation of both data of the atmospheric neutrino 
and the solar neutrino by the neutrino mass matrix derived from the
mixing of neutrinos and gauginos in the extended MSSM with an
extra U(1) gauge symmetry. 
This scenario directly  relates the neutrino mass to supersymmetry.
The structure of the mass hierarchy and the mixing is determined 
only by the extra U(1)-charge of neutrinos.
Although the model is rather simple, it may be able to realize 
both the small and large mixing angle solutions for the solar neutrino
problem.
In particular, the large mixing angle MSW solution for the solar
neutrino problem can be easily realized.  
We discuss the relation between the extra U(1)-charge and four
solutions for the solar neutrino problem.

\newpage
\setcounter{footnote}{0}
\def\thefootnote{\arabic{footnote}}

The atmospheric and solar neutrino observations have strongly indicated
the existence of the neutrino oscillation and then the non-zero neutrino
mass \cite{oscil1,oscil2,sk}. 
This is an only known evidence suggesting that there is 
new physics beyond the standard model (SM). 
Theoretically, the SM has been claimed
that there are some unsatisfactory features. Such a representative problem is
a gauge hierarchy problem. Supersymmetry has been considered to be a
promising candidate to solve it and its various phenomenological
features have been studied almost for twenty years \cite{super}. 
However, still now 
there seems not to be an experimental support for it, except for the 
gauge coupling unification \cite{unify}. 
If we can find that there are some relations 
between the supersymmetry and the neutrino mass, it is very 
interesting and might also cause a big impact to the study of the 
new physics.

In general the supersymmetric model has a special symmetry called
$R$-parity defined by $R_p=(-1)^{3B+L+2S}$, where $B$ is the baryon
number, $L$ is the lepton number and $S$ is the spin.
The ordinary particles in the SM are $R_p$-even and their
superpartners are $R_p$-odd. As far as the $R$-parity is conserved,
neutrinos cannot mix with neutralinos.
However, the superpotential can include an $R$-parity violating
term like $\epsilon_\alpha L_\alpha H_2$ \cite{rparity1,rparity2}, 
for example, 
as a result of the gravitational effect.
Under the existence of this bi-linear $R$-parity violating term, 
the vacuum expectation values (VEVs) of sneutrinos is generated and then
the mixing among neutrinos and neutralinos is induced.
The generation of Majorana neutrino mass in the minimal supersymmetric SM
(MSSM) through this mixing has been discussed in a lot of works 
\cite{rmass1,rparity1,rparity2}.
In most of them only one of three neutrinos is massive at the tree
level \cite{rmass2} and then they could be applied only to the atmospheric 
neutrino problem \cite{treeatm}. 
It is necessary to improve this feature in order to 
explain all of the recent neutrino oscillation data\footnote{In this paper we 
confine our study into the explanation of the atmospheric and solar 
neutrino data. 
It might be straightforward to extend the model to include a sterile neutrino 
and make it applicable to the LSND data \cite{lsnd}.}. 
One of the possibilities to make other neutrinos massive is 
to introduce the one-loop effect \cite{loop,ratm}.  
In such a framework the solar and atmospheric neutrino problems have
been also discussed \cite{ratm}.   

In this paper we would like to propose the generation of the 
small neutrino mass at the tree level based on a typical interaction among
neutrinos and gauginos in the usual supersymmetric model. 
The model is very simple and economical. It needs only a small
extension of the MSSM by an extra U(1) gauge symmetry.
There are some works \cite{exsym} on the phenomenology including the
neutrino mass generation in the supersymmetric model with the
spontaneous $R$-parity violation and the extra gauge symmetry.  
A new feature of our model is that an introduced U(1)-symmetry has
generation dependence. 
The neutrino mass matrix produced in this way has noticeable 
features for the explanation of both the atmospheric and solar
neutrino data.
It might realize the large mixing angle solution  
for them in a natural way if we can assume the spontaneous lepton number 
violation due to the small VEVs of sneutrinos.
Our model could be distinguished at least by the existence of the additional
neutral gauge boson $Z^\prime$ at the TeV region from other 
neutrino models with the large mixing solutions for the solar neutrino
problem \cite{bimaxim,bimaxim2}.

In the MSSM there are two neutral gauginos $\lambda_{W_3}$ and $\lambda_Y$
which are the superpartners of the SU(2)$_L$ and U(1)$_Y$ gauge fields.
Their interaction with the ordinary left-handed neutrinos 
$\nu_\alpha~(\alpha=e,\mu,\tau)$ can be written as \cite{super}
\begin{equation}
{i\over \sqrt 2}g_2\sum_\alpha\left(\tilde\nu_\alpha^\ast
\lambda_{W_3}\nu_\alpha 
-\bar\lambda_{W_3}\bar\nu_\alpha\tilde\nu_\alpha\right)
-{i\over \sqrt 2}g_1\sum_\alpha\left(\tilde\nu_\alpha^\ast
\lambda_Y\nu_\alpha 
-\bar\lambda_Y\bar\nu_\alpha\tilde\nu_\alpha\right).
\label{eqa}
\end{equation}
If the spontaneous R-parity violation occurs by the VEVs of sneutrinos 
$\langle\tilde\nu_\alpha\rangle\not=0$, the mixing among gauginos 
and neutrinos is induced. On the other hand, the supersymmetry breaking
causes the gaugino masses $M_2$ and $M_1$ for $\lambda_{W_3}$ and
$\lambda_Y$ through a suitable mechanism.
In the case of $M_{1,2} \gg g_{1,2}\langle\tilde\nu_\alpha\rangle$, 
we can expect to obtain the small Majorana masses of neutrinos 
as a result of a kind of seesaw mechanism. 
In fact, if we assume 
$\langle\tilde\nu_e\rangle=\langle\tilde\nu_\mu\rangle
=\langle\tilde\nu_\tau\rangle=u$, the mass mixing among neutrinos and
gauginos can be expressed as 
${\cal L}_{\rm mass}=-{1\over 2}({\cal N}^T{\cal M}{\cal N}+{\rm h.c.})$ and
\begin{equation}
{\cal M}=\left(\begin{array}{ccccc}
0&0&0&{g_2\over\sqrt 2}u & {g_1\over\sqrt 2}u\\ 
0&0&0&{g_2\over\sqrt 2}u & {g_1\over\sqrt 2}u\\ 
0&0&0&{g_2\over\sqrt 2}u & {g_1\over\sqrt 2}u\\ 
{g_2\over\sqrt 2}u & {g_2\over\sqrt 2}u &{g_2\over\sqrt 2}u&M_2&0 \\
{g_1\over\sqrt 2}u & {g_1\over\sqrt 2}u &{g_1\over\sqrt 2}u& 0&M_1\\
\end{array}\right),
\label{eqb}
\end{equation}
where ${\cal N}^T=(\nu_\alpha, -i\lambda_{W_3},i\lambda_Y)$.
We consider a special case in which every VEV of sneutrinos is equal 
in order to reduce the number of parameters and simplify the 
mass matrix of the model.  
Under the assumption of $M_{1,2}\gg g_{1,2}u$ we can easily diagonalize 
the mass matrix ${\cal M}$. Although there are three light mass eigenvalues,
two of them are zero\footnote{Even if we do not take the VEVs of 
sneutrinos equal, the situation is the same. In order to improve it the
one-loop effect has been taken into account in ref.~\cite{rparity2,ratm}.}. 
We must extend the MSSM to apply this type
of mass generation of neutrinos to the explanation of the atmospheric and
solar neutrino data. We consider a model which can naturally change that
aspect at the tree level.

\def\romth{I\hspace*{-0.8mm}I\hspace*{-0.8mm}I}
\def\romtw{I\hspace*{-0.7mm}I}
Let us consider the introduction of an extra U(1)$_X$ gauge symmetry specified 
by the following features\footnote{The additional U(1)-symmetry is known 
to appear very often in the heterotic superstring models. 
It can play some useful roles in the supersymmetric models \cite{z00}.}.
It is non-anomalous and also has flavor diagonal but non-universal 
couplings. 
The different charge of U(1)$_X$ is assigned to the fields belonging 
to the different generation\footnote{This 
kind of charge assignment of U(1)$_X$ has been discussed in the different
scenario to explain the small neutrino mass and the proton stability, for
example, in \cite{sue2}.}.
It is assumed to be spontaneously broken at a TeV scale region and then
we have an additional neutral gauge bozon $Z^\prime$ which is heavier 
than the ordinary $Z^0$.
We need a new SM singlet chiral superfield whose scalar component 
causes the spontaneous breaking of U(1)$_X$ by its VEV.  
Since a part of the MSSM contents is assumed to have its 
charge in a generation dependent way, it could generate a 
non-trivial texture in the mass matrices.
However, we cannot use this symmetry for the Froggatt-Nielsen 
mechanism \cite{fn} to induce the hierarchical structure of 
quark mass matrices since its breaking scale is too small as 
compared to the Planck scale. 
For simplicity, we do not assign the U(1)$_X$-charge to quarks
and we consider a model in which only leptons have its charges as
\begin{equation}
\ell_L \quad (q_I, ~q_I, ~q_{\romth}), \qquad 
\bar \ell_R \quad (-q_I, ~-q_I, ~-q_{\romth}).
\label{charge}
\end{equation}
As a result of such a charge assignment $\nu_{_{\romth}}$ has a 
charge $q_{_{\romth}}$ different from other neutrinos $\nu_I$ and 
$\nu_{\romtw}$ whose charge 
is defined by $q_{I}$\footnote{At this stage we cannot 
determine to which flavor each $\nu_\alpha $ corresponds so that we used 
the Roman numerals for $\alpha$.}. 
The doublet Higgs superfields are also assumed to have no charge 
so that the ordinary
Yukawa couplings of the charged leptons induce the texture in 
the mass matrix such as $m_{_{I\romth}}=m_{_{\romtw\romth}}=0$.
Using this charge, the coupling between neutrinos and the 
U(1)$_X$ gaugino is given by
$i\sqrt 2g_{_X}\sum_\alpha q_\alpha\left(\tilde\nu_\alpha^\ast
\lambda_X\nu_\alpha -\bar\lambda_X\bar\nu_\alpha\tilde\nu_\alpha\right)$
as eq.~(\ref{eqa}).
We do not consider the kinetic term mixing between the U(1)-gauginos
\cite{sue1}. If we take this effect into account, 
off-diagonal elements appear in the gaugino mass matrix.
In general, the introduction of U(1)$_X$ to the MSSM requires addtional
chiral superfields to cancel the gauge anomaly which causes the non-trivial
constraint on the charge assignment. Since its breaking scale is in 
the TeV region, the non-universal couplings of $Z^\prime$ with leptons
might impose the constraints on the model through the electroweak precision 
measurements and the flavor changing neutral current (FCNC). 
We will discuss these points later.

First we focus our attention to the role of U(1)$_X$ in the generation of
the neutrino mass due to the mixing among neutrinos and gauginos.
Since the gaugino sector is extended, the matrix (\ref{eqb}) is 
modified into\footnote{In the
phenomenological point of view the result obtained in this paper 
using this mass matrix is independent of the mass generation
mechanism. If we can find any models giving the neutrino mass matrix 
with the features presented here, 
the result could be also applied to them.}
\begin{equation}
{\cal M}=\left(\begin{array}{cc}0 &m^T \\ m & M \\ \end{array} \right),
\qquad
m=\left(\begin{array}{ccc}
a_2 & a_1 & b \\ a_2 & a_1 & b \\ a_2 & a_1 & c \\
\end{array}\right), 
\qquad
M=\left(\begin{array}{ccc}
M_2 & 0 & 0 \\ 0 & M_1 & 0 \\ 0 & 0 & M_X \\
\end{array}\right),
\label{eqc}
\end{equation}
where $a_{\ell}={g_{\ell}\over\sqrt 2}u$, $b=\sqrt 2g_{_X}q_{_I}u$ and 
$c=\sqrt 2g_{_X}q_{_{\romth}} u$.  
We can obtain the light neutrino mass matrix from this by
using the generalized seesaw formula and it can be written as
\begin{equation}
M_\nu = m^T M^{-1}m =\left(\begin{array}{ccc}
m_0+\epsilon^2 & m_0+\epsilon^2 & m_0+\epsilon\delta \\
m_0+\epsilon^2 & m_0+\epsilon^2 & m_0+\epsilon\delta \\
m_0+\epsilon\delta & m_0+\epsilon\delta & m_0+\delta^2 \\
\end{array}\right),
\label{eqd}
\end{equation}
where $m_0,\epsilon$ and $\delta$ are defined by 
\begin{equation}
m_0={g_2^2u^2\over 2M_2}+{g_1^2u^2\over 2M_1}, \qquad 
\epsilon={g_{_X}q_{_I}u\over\sqrt{M_X}}, \qquad
\delta={g_{_X}q_{_{\romth}}u\over\sqrt{M_X}}.   
\end{equation}
The interesting aspect of this mass matrix is that it is defined only by the
 gaugino mass $M_A~(A=\ell, X)$, the gauge couplings $g_{_A}$, the 
U(1)$_X$-charges $q_\alpha$ and the VEV $u$ of sneutrinos. 

We define the mass eigenstates by 
$\tilde{\cal N}_i=(U^T)_{i\alpha}{\cal N}_\alpha$. 
The diagonalization of the matrix (\ref{eqd}) gives
\begin{equation}
U^T=\left(\begin{array}{ccc}
{1\over\sqrt 2} & -{1\over\sqrt 2} & 0 \\
{\cos\theta\over\sqrt 2} & {\cos\theta\over\sqrt 2} & -\sin\theta \\
{\sin\theta\over\sqrt 2} & {\sin\theta\over\sqrt 2} & \cos\theta \\
\end{array}\right),
\label{eqe}
\end{equation}
where one of the mixing angles $\sin\theta$ is defined as
\begin{equation}
\sin^22\theta ={8(m_0+\epsilon\delta)^2\over (m_0+2\epsilon^2-\delta^2)^2
+8(m_0+\epsilon\delta)^2}.
\label{eqf}
\end{equation}
The non-zero mass eigenvalues are
\begin{equation}
m_{2,3}={1\over 2}\left\{(3m_0+2\epsilon^2+\delta^2)
\mp\sqrt{(m_0+2\epsilon^2-\delta^2)^2+8(m_0+\epsilon\delta)^2}\right\}.
\label{eqg}
\end{equation}
Here we can consistently assume the charged lepton mass matrix is
diagonal\footnote{ 
Although the U(1)$_X$ constrains the texture of charged 
leptons mass matrix as mentioned before, some additional symmetry might be
necessary to make it diagonal. The large mixing between $\mu$ and $\tau$ 
is disfavored in this model.}.
In this case the above mixing matrix $U$ is just the
flavor mixing matrix which controls the neutrino oscillation. 
In the following discussion we assume it in the charged lepton sector.
There is no $CP$ violation in the lepton sector under this assumption. 

\begin{figure}[tb]
\begin{center}
\begin{tabular}{c|ccc}\hline
$(\alpha,\beta)$ & $(i,j)$ & 
$-4U_{\alpha i}U_{\beta i}U_{\alpha j}U_{\beta j}(\equiv{\cal A})$  & \\ \hline\hline
$(I,\romtw)$& $(1,2)$& $\cos^2\theta$  & (A)\\
                    & $(1,3)$& $\sin^2\theta$ & (B)\\
                    & $(2,3)$& $-\sin^2\theta\cos^2\theta$ & (C)\\ \hline
$(I, \romth)$ & $(2,3)$& $2\sin^2\theta\cos^2\theta$ & (D) \\ \hline
$(\romtw, \romth)$ & $(2,3)$& $2\sin^2\theta\cos^2\theta$ & (E)\\ \hline
\end{tabular}
\end{center} 
\vspace*{1mm}

{\footnotesize Table 1.~~ The contributions to each neutrino transition
process $\nu_\alpha\rightarrow\nu_\beta$ from each sector 
$(i, j)$ of the mass eigenstates.}
\end{figure}

It is well-known that the transition probability due to the
neutrino oscillation  
$\nu_\alpha\rightarrow \nu_\beta$ after the flight length $L$ is written 
by using the matrix elements of (\ref{eqe}) as
\begin{equation}
{\cal P}_{\nu_\alpha\rightarrow \nu_\beta}(L)
=\delta_{\alpha\beta}
-4\sum_{i>j}U_{\alpha i}U_{\beta i}U_{\alpha j}U_{\beta j}
\sin^2\left({\Delta m_{ij}^2\over 4E}L\right),
\label{eqh}
\end{equation}
where $\Delta m_{ij}^2=\vert m_i^2-m_j^2\vert$.
In Table 1 we summalize the contribution to each neutrino transition
mode $(\alpha, \beta)$ from each sector $(i, j)$ of the 
mass eigenstates in our model. 
As a phenomenologically interesting case, we consider the 
case in which the mass eigenstates $\tilde\nu_2$ and $\tilde\nu_3$ are almost 
degenerate and the mass hierarchy
$m_1 \ll m_2 \simeq m_3$ is satisfied\footnote{
This is a well-known reversed hierarchy scenario to the atmosphetic and solar
neutrino problems \cite{invert}. However, the absolute value of each
mass eigenvalue is smaller than the usual scenario because of $m_1=0$.
Every neutrino cannot be a hot dark matter candidate.}.
If we use it to explain the atmospheric and solar neutrino data, 
the difference of the squared mass should be taken as 
\begin{eqnarray}
&&2\times 10^{-3}~{\rm eV}^2~{^<_\sim}~ 
\Delta m^2_{12}\simeq \Delta m^2_{13}~{^<_\sim}~ 6\times 10^{-3}~{\rm eV}^2, 
\label{eqha}\\
&&10^{-10}~{\rm eV}^2 ~{^<_\sim}~ \Delta m^2_{23}~{^<_\sim}~ 
1.5\times 10^{-4}~{\rm eV}^2.
\label{eqhb}
\end{eqnarray} 
The suitable value of $\Delta m^2_{23}$ should be chosen within the
above range depending on which
solution is adopted for the solar neutrino problem.
We can easily find the simultaneous explanation of
both deficit of the atmospheric neutrino and the solar neutrino if we identify
the weak eigenstates $(\tau,\mu,e)$ with $(I, \romtw, \romth)$.
Using this identification, eq.~(\ref{eqe}) is rewritten into the 
MNS mixing matrix in the usual basis as, 
\begin{equation}
U^{\rm MNS}=\left(\begin{array}{ccc}
0 &-\sin\theta & \cos\theta \\
-{1\over\sqrt 2} & {\cos\theta\over\sqrt 2} & {\sin\theta\over\sqrt 2} \\
{1 \over\sqrt 2} & {\cos\theta\over\sqrt 2} & {\sin\theta\over\sqrt 2} \\
\end{array}\right).
\label{eqhc}
\end{equation}
The atmospheric neutrino is found to be explained by
$\nu_\mu\rightarrow\nu_\tau$ composed of (A) and (B) in Table 1.
Here we should note that $m_1=0$ and also
$\Delta m_{12}^2\simeq\Delta m_{13}^2$ is satisfied.
This explanation is independent of the value of $\sin\theta$. 
On the other hand, the solar neutrino is expected to be explained by
$\nu_e\rightarrow\nu_\mu$ (E) and also 
$\nu_e\rightarrow\nu_\tau$ (D).
In both processes the amplitude ${\cal A}$ is ${1\over 2}\sin^22\theta$.
If $\sin^22\theta\sim 10^{-2}$, it could realize the MSW small mixing angle
solution (SMA) \cite{sma}. 
If $\sin^22\theta\sim 1$, it could give the 
MSW large mixing angle solution (LMA), the low mass MSW solution (LOW)
 and the vacuum oscillation solution (VO) \cite{sma} depending on the
value of $\Delta m_{23}^2$.
The CHOOZ experiment \cite{chooz} constrains a component $U_{e1}$ 
of the MNS mixing matrix in this scenario 
since the amplitude ${\cal A}$ of the contribution to 
$\nu_e \rightarrow \nu_x$ with $\Delta m_{12}^2$ or $\Delta m_{13}^2$
always contains it. This model is free from this constraint since
$U_{e1}=0$.  

\input epsf 
\begin{figure}[tb]
\begin{center}
\epsfxsize=7.6cm
\leavevmode
\epsfbox{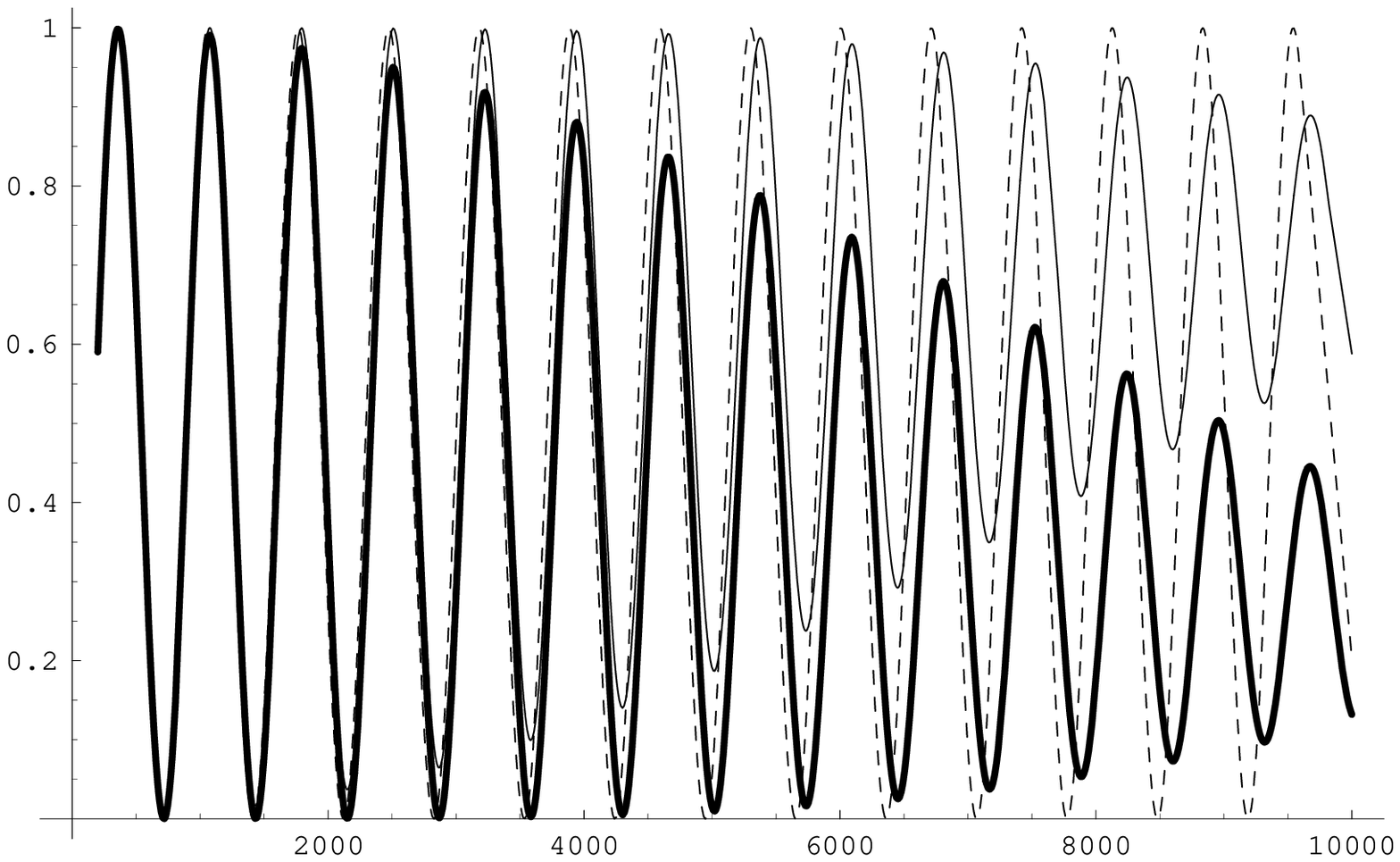}
\end{center}
\vspace*{-0.6cm}
{\footnotesize Fig. 1~~\  The transition probability
 $P(\nu_\mu\rightarrow\nu_{x(\not=\mu)})$ as a function of 
the flight length $L$. 
We assume $E=$1~GeV, $\Delta m_{13}^2=3.5\times 10^{-3}$~eV$^2$ and
$\Delta m_{23}^2= 10^{-4}$~eV$^2$}
\end{figure}

In Table. 1 an only remaining contribution (C) to 
$\nu_\mu \rightarrow \nu_\tau$
cannot imply any evidence in the short baseline experiment even in the
case of $\sin^22\theta\simeq 1$ since $\Delta m_{23}^2$ is too small. 
However, this mode might be relevant to the long baseline experiment 
in the case of 
$\Delta m_{23}^2\sim 10^{-4}$~eV$^2$ which corresponds to the LMA
solution of the solar neutrino deficit. We show the effect of the
mode (C) on the $P(\nu_{\mu} \rightarrow \nu_x)$ in Fig. 1.
The dashed line comes from the modes (A) and (B). This corresponds
to the ordinary two flavor oscillation. 
The thick solid line is the one which is obtained by 
taking account of the contribution of
(C). In the thin solid line the contribution of (D) which corresponds
to $\nu_x=\nu_e$ is also taken into account.
This shows that it might be possible to discriminate this model from 
others in the long baseline experiment such as $L~{^>_\sim}~2000$~km. 
Moreover, this model may be expected to have another experimental signature 
in the neutrinoless double $\beta$-decay \cite{bb0}.
Using eq.~(\ref{eqhc}), the effective mass parameter which appears in a
formula of the rate of neutrinoless double $\beta$-decay 
is estimated as
\begin{equation}
\vert m_{ee}\vert=\left\vert \sum_j\vert U_{ej}
\vert^2e^{i\phi_j}m_j\right\vert
=\left(m_2\sin^2\theta+m_3\cos^2\theta\right)\sim m_3.
\end{equation}
Thus $\vert m_{ee}\vert$ takes the value in the range of 0.04 - 0.08~eV 
which is independent of the value of $\sin\theta$, that is, the solution 
of the solar neutrino problem.
The value seems to be within the reach in the near future experiment.

In order to see whether the values of the oscillation parameters
can be successfully realized, we need to study 
them by using the parameters $M_A$, $g_{_A}$, $q_\alpha$ and $u$.
In the usual soft supersymmetry breaking scenario the gaugino mass
is universally produced as $M_0$ at the unification scale. 
Its low energy value is determined by the 
renormalization group equations (RGEs).
If we use the one-loop RGEs, the gaugino mass at a scale $\mu$ 
can be expressed as
\begin{equation}
M_2(\mu) = {M_0\over g_U^2}g_2^2(\mu), \qquad
M_1(\mu) = {5\over 3}{M_0\over g_U^2}g_1^2(\mu),
\label{eqi}
\end{equation}
where we assume the gauge coupling unification at the scale $M_U$ 
and define a value of the gauge coupling at $M_U$ as $g_{_U}$. 
It is not unnatural to
consider the gauge coupling of U(1)$_X$ and its gaugino mass to be the 
same as the ones of U(1)$_Y$ \footnote{It might be 
satisfied if U(1)$_X$ is unified into a simple group together 
with other SM gauge group like the SO(10) 
and E$_6$ models. Also in the superstring context the freedom of an Abelian 
Kac-Moody level could make it possible.}.
If we adopt this simplified possibility, we can find that $m_{2,3}$ and
$\sin^22\theta$ can be written by 
using only $q_\alpha$ and ${g_U^2\over M_0}u^2$ as
\begin{eqnarray}
&&m_{2,3}={3\over 5}\left(2+2q_{_I}^2+q_{_{\romth}}^2 \mp
\sqrt{(2+2q_{_I}^2+q_{_{\romth}}^2)^2-{16\over 3}
(q_{_I}-q_{_{\romth}})^2}\right){g_U^2u^2\over M_0}, \label{eqj} \\
&&\sin^22\theta={8(2+3q_{_I}q_{_{\romth}})^2\over
(2+6q_{_I}^2-3q_{_{\romth}}^2)^2 +8(2+3q_{_I}q_{_{\romth}})^2}.
\label{eqk}
\end{eqnarray}
The structure of the mass spectrum and the flavor mixing
is controled by the U(1)$_X$-charge. The gaugino mass $M_0$ and the
VEV $u$ of sneutrinos are relevant to the 
mass eigenvalues only in the form of an overall factor ${g_U^2\over M_0}u^2$.
In order to realize the value of (\ref{eqha}) which is
required by the atmospheric neutrino deficit, ${g_U^2\over M_0}u^2$ 
needs to be in the range of 0.017 eV~${^<_\sim}~{g_U^2\over M_0}u^2~
{^<_\sim}~0.023$ eV~ \footnote{In the estimation we have already taken
account of the effect coming from the U(1)$_X$-charge dependence. However, its 
effect is not large and only a factor of $O(1)$.}.
If we take $M_0\sim 100$~GeV and $g_{_U}\sim$0.72, for example,
it shows that the sneutrino VEV should be $u\sim 60$ - 70~keV.
The remaining freedom which we can use to explain the solar neutrino
deficit is the U(1)$_X$-charge of neutrinos.  
In Fig. 2 we plot the value of the U(1)$_X$-charge and the corresponding 
oscillation parameters which are suitable for the explanation of the
solar neutrino deficit.
The figure shows that the reasonable value of the U(1)$_X$-charge
can realize all the well-known solutions for the solar neutrino problem.
As is easily seen from eq.~(\ref{eqj}), 
$\Delta m^2_{23}$ is propotional to the part of square root in the
expression of $m_{2,3}$.
It means that the solution with the smaller $\Delta m^2_{23}$ requires
the finely tuned U(1)$_X$-charge\footnote{It may be useful to
note that the complete degeneracy $m_2=m_3$ occurs in the case of
$q_{_I}=1/\sqrt 3$ and $q_{_{\romth}}=-2/\sqrt 3$, which satisfy 
$\sum_\alpha q_\alpha=0$.}.
As a result, the SMA, the LOW and the VO seem to require the 
finer tuning of the U(1)$_X$-charge than the LMA.
It might be favorable for this model if we take seriously
the recent Super-Kamiokande analysis of the solar neutrino 
\cite{sk}.  
Anyway it is interesting that every solution is obtainable within this
simplified framework. 

\input epsf 
\begin{figure}[tb]
\begin{center}
\epsfxsize=7.6cm
\leavevmode
\epsfbox{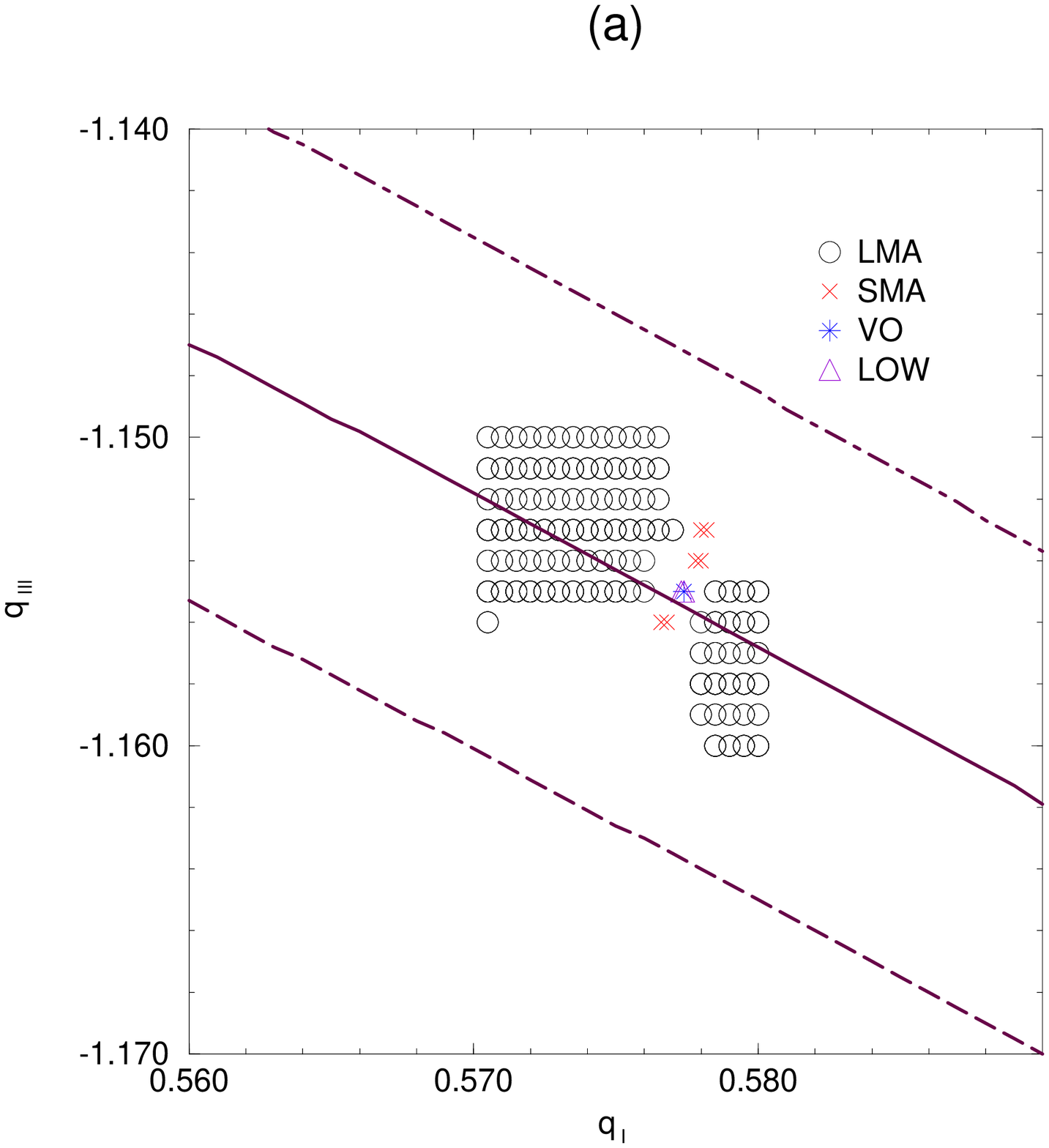}
\hspace*{0.3cm}
\epsfxsize=7.6cm
\leavevmode
\epsfbox{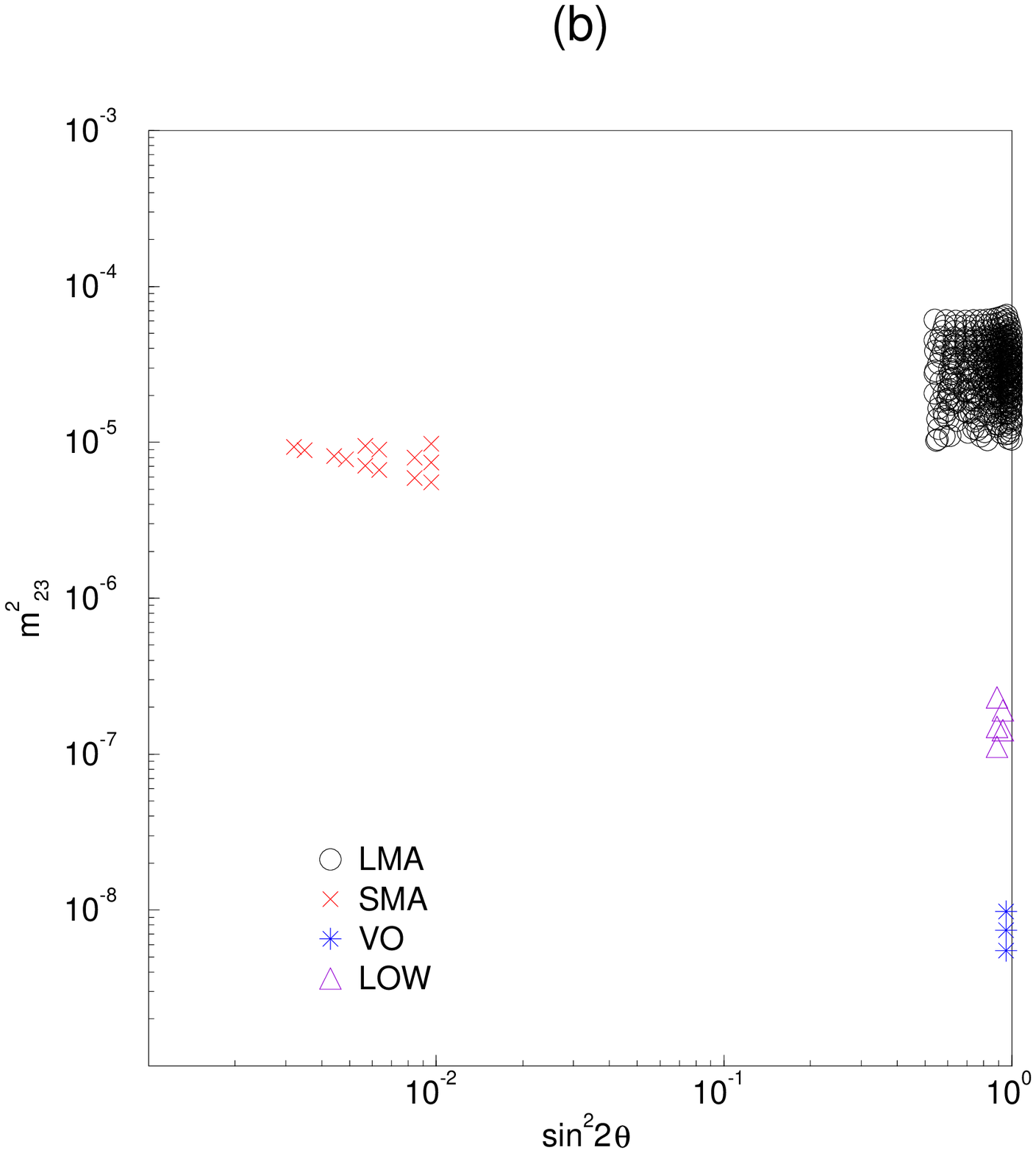}
\end{center}
\vspace*{-1cm}
{\footnotesize Fig. 2~~\  (a) Scatter plots of the U(1)$_X$-charge
of neutrinos which can give the solution to the atmospheric and solar
neutrino problems. The dot-dashed, solid and dashed lines are the ones 
corresponding to $Q=1.04,~1.05$ and 1.06, respectively.
(b) Oscillation parameters for the solar neutrino
which are realized by the U(1)$_X$-charge of neutrinos shown in (a).}
\end{figure}

Now we comment on some remaining important points.
The first point is whether the anomaly-free U(1)$_X$ can realize the
solutions in Fig.~2(a). We show it by using a simple example presented in
(\ref{charge}). In order to cancel the gauge anomaly
 we introduce additional chiral superfields with the non-zero U(1)$_X$-charge: 
${\bf 2}_0,~{\bf 1}_{\pm 1}$ and $4({\bf 1}_0)$, where we 
indicate the representation under SU(2)$_L\times$U(1)$_Y$.
The U(1)$_X$-symmetry is assumed to be broken by the VEVs of ${\bf 1}_0$'s. 
The ${\bf 2}_0$ and ${\bf 1}_{\pm 1}$ should be massive at the TeV scale 
through the VEVs. If we impose this requirement 
with the anomaly-free conditions 
on the U(1)$_X$-charges, we find that the charges of all fields can 
be expressed by using the charge $Q$ of one of ${\bf 1}_0$'s. 
In Fig.~2(a) we plot the values of $(q_I, q_{\romth})$ of such 
solutions by taking suitable values of $Q$. 
From it we find that our scenario could have the solutions consistent
with the above mentioned conditions. 
Although we take a very simple example to show it, it would be 
possible to construct more elaborate examples. 
 
The next point is the phenomenological constraint on the U(1)$_X$ since
we have a $Z^\prime$ at a scale not far from the weak scale.
The constraint on the $Z^\prime$ comes from the direct search
\cite{direct} and the electroweak precision measurements \cite{z0,z00}. 
In general these seem to require the mass bound $m_{Z^\prime}>600$~GeV, 
which can be consistent with our assumption of the U(1)$_X$ breaking scale.
Another constraint could also come from the FCNC in the lepton sector such as 
the coherent $\mu$-$e$ conversion, $\tau\rightarrow 3e,~3\mu$ and 
$\mu\rightarrow e\gamma$ {\it etc.}, since we assume the flavor diagonal
but non-universal couplings of $Z^\prime$ to leptons. 
The detailed analysis for such a issue has been done in \cite{fcnc} 
and we can use the discussion there. In our model the $Z^\prime$ 
interaction term can be written in the mass eigenstates as,
\begin{eqnarray}
&&{\cal L}_{Z^\prime}=-g_1\left[{g_X\over g_1}\cos\xi J_{(X)}^\mu
-\sin\xi J_{(1)}^\mu\right] Z_\mu^\prime, \nonumber \\
&&J_{(X)}^\mu=\sum_{ij}\left[\bar\nu_{L_i}B_{ij}^{\nu_L}\nu_{L_j}
+\bar\ell_{L_i}B_{ij}^{\ell_L}\ell_{L_j}+\bar 
\ell_{R_i}B_{ij}^{\ell_R}\ell_{R_j}\right], \nonumber \\ 
&&B_{ij}^\psi= V^{\psi\dagger}~ {\rm diag}(q_{\romth}, q_I, q_I)~V^\psi,
\end{eqnarray}
where $\xi$ is a $Z$-$Z^\prime$ mixing angle and $V^\psi$ is the
unitary matrix to diagonalize the mass matrix of $\psi$.
The FCNC in the lepton sector causes the strong constraint on
$B_{ij}^{\ell_{L,R}}$ as it is discussed in \cite{fcnc}.
Since the U(1)$_X$-charge of $\mu$ and $\tau$ is equal in our model, 
$B_{23}^{\ell_{L,R}}=0$.
Moreover, if we remind that $m_{e\mu}=m_{e\tau}=0$ is satisfied in
the charged lepton mass matrix because of the constraint of 
the U(1)$_X$ charge,
we find that $B_{12}^{\ell_{L,R}}=B_{13}^{\ell_{L,R}}=0$.
Then the non-universal couplings of $Z^\prime$ do not induce the
observable effect on the FCNC.
The addtional contribution to the FCNC could be caused by the extended 
neutralino sector. The study in that case has been done
in \cite{sue1} for the models which can be embedded in E$_6$.
The effect is found not to be so large as far as we take parameters
in such a way that the MSSM satisfies the experimental bound.
Although the U(1)$_X$-symmetry is different from this case, 
the qualitative feature is similar also in the present model.
Thus as far as we consider a TeV scale $Z^\prime$ with the small 
mixing $\xi~{^<_\sim}~10^{-3}$, the constraint from the precision
measurement and the FCNC of the lepton sector seems not severe to 
the present model. 

Another important point is the origin of the small VEV $u$ of sneutrinos.
As mentioned in the previous part, it should be around $O(10^2)$~keV
which is much smaller than the weak scale.
In the MSSM there are arguements on the lepton number violation 
due to the VEVs of sneutrinos in the vicinity of the weak scale \cite{rviol} 
and also some authors point out that the neutrino mass produced by 
them can be sufficiently small \cite{ratm,rparity2}.
However, in our scenario we need much smaller VEVs of sneutrinos than
the weak scale. 
As one of the possibilities, we might consider that
such small VEVs of sneutrinos could be obtained if there is
bi-linear $R$-parity violating terms $\epsilon L_\alpha H_2$
with the sufficiently small $\epsilon$ \footnote{
We assume $\epsilon_i=\epsilon$. The smallness of
$\epsilon$ also causes a new hierarchy problem. To resolve this problem
we might be able to use the similar mechanism to the solutions 
for the $\mu$ problem \cite{muprob}, 
although the relevant energy scale needs to be scaled down by some orders 
of magnitude.}. 
We can check it briefly by minimizing the scalar
potential under the assumption that $\langle H_{1,2}\rangle$ can be treated 
as constants. 
In this case the value of $u$ derived from the potential minimization 
condition can be approximately written as
\begin{equation}
u \sim {6\epsilon(\mu \langle H_1\rangle+B_\epsilon\langle H_2\rangle )\over
3(g_1^2+g_2^2)(\langle H_1\rangle^2-\langle H_2\rangle^2)+2g_X^2(\sum_
\alpha q_\alpha)(q_1\langle H_1\rangle^2+q_2\langle H_2\rangle^2)+6m^2},
\end{equation}
where $B_\epsilon$ is a soft supersymmetry breaking parameter 
related to the $\epsilon L_\alpha H_2$ terms and $m^2$ is the soft 
scalar mass of
sneutrinos, which is assumed to be universal. 
From this we find that the sufficiently small $u$ can be obtained as far 
as $\epsilon$ is small enough and the $\mu$-parameter, $B_\epsilon$ 
and $m^2$ take the values of the order of weak scale. 
However, the $\epsilon L_\alpha H_2$ terms generate the $\nu_\alpha$-$\tilde
H_2^0$ mixing which affects our analysis.
In order to escape the influence due to the $\epsilon$ terms, 
${g_{_A} u\over M_A}$ might be required to be sufficiently larger than 
${\epsilon\over \mu}$. 
We need to check whether these conditions are satisfied  
at the true vacuum with the equal VEVs of sneutrinos taking account 
of the radiative correction. 
Including the study of other possibility to realize the small $u$,
we would like to leave it in the future publication 
instead of going further into it here. 

In summary we have proposed the scenario for the mass and the mixing of
neutrinos in the supersymmetric model with an extra U(1)$_X$-symmetry.
The scenario is based on the mixing among neutrinos and gauginos. 
In this model we could obtain 
the non-zero mass eigenvalues for neutrinos at the tree level. 
The atmospheric and solar neutrino deficits can be simultaneously 
explained by the reversed mass hierarchy scenario. 
In particular, every known solution for the solar
neutrino problem can be realized only by tuning the U(1)$_X$-charge of 
neutrinos. It is interesting that the large mixing angle MSW 
solution can be easily realized as compared to other solutions.
The neutrinoless double $\beta$-decay might be accessible if the
experimental bound is improved to the level of $\vert m_{ee}\vert\sim$~0.04
- 0.08~eV.
There remain some unsolved problems related to the VEVs of sneutrinos
and the supersymmetry breaking parameters. 
Further investigation of these problems seems to be necessary 
to see whether our model works well in a realistic way. 
Finally we would like to stress that the texture of the present mass
matrix might be scaled up into the high energy region. 
In that case it may be
regarded as the usual seesaw model with a special Dirac mass matrix.
Various conditions on the U(1)$_X$-charge {\it etc.} in the present
scenario could be replaced into the ones for Yukawa couplings and so on. 
The study in this direction also seems to be interesting. 
\vspace*{3mm}

\noindent
I would like to thank M.~Tanimoto for useful comments related to
the CHOOZ constraint.
This work is supported in part by the Grant-in-Aid for Scientific 
Research from the Ministry of Education, Science and Culture
(No.11640267).

\newpage

\end{document}